\documentclass[showpacs,floatfix,superscriptaddress,twocolumn,amssymb,amsfonts,prl,aps]{revtex4}
\usepackage{longtable,graphicx,color,epsfig,dcolumn}
\begin{document}
\bibliographystyle{revtex}
\title
{Atomic-scale surface demixing in a eutectic liquid BiSn alloy}
\author{Oleg~G.~Shpyrko}
\email{oleg@xray.harvard.edu} \altaffiliation[present
address:]{Center for Nanoscale Materials, Argonne National
Laboratory, Argonne, IL, 60439} \affiliation{Department of Physics
and DEAS, Harvard University, Cambridge MA 02138}

\author{Alexei~Yu.~Grigoriev}
\affiliation{Department of Physics and DEAS, Harvard University,
Cambridge MA 02138}

\author{Reinhard Streitel}
\affiliation{Department of Physics and DEAS, Harvard University,
Cambridge MA 02138}

\author{Diego Pontoni}
\affiliation{Department of Physics and DEAS, Harvard University,
Cambridge MA 02138}

\author{Peter~S.~Pershan}
\affiliation{Department of Physics and DEAS, Harvard University,
Cambridge MA 02138}

\author{Moshe~Deutsch}
\affiliation{Department of Physics, Bar-Ilan University, Ramat-Gan
52900, Israel}

\author{Ben~Ocko}
\affiliation{Department of Physics, Brookhaven National Laboratory,
Upton NY 11973}

\author{Mati Meron}
\affiliation{CARS, University of Chicago, Chicago, IL 60637}

\author{Binhua Lin}
\affiliation{CARS, University of Chicago, Chicago, IL 60637}

\date{\today}

\begin{abstract}
\def\baselinestretch{1}
\keywords{demixing, phase separation, Gibbs adsorption, liquid
alloys, binary mixtures, bismuth, tin, surface, segregation, x-ray,
resonant, reflectivity, synchrotron}

\noindent Resonant x-ray reflectivity of the surface of the liquid
phase of the Bi$_{43}$Sn$_{57}$ eutectic alloy reveals atomic-scale
demixing extending over three near-surface atomic layers. Due to the
absence of underlying atomic lattice which typically defines
adsorption in crystalline alloys, studies of adsorption in liquid
alloys provide unique insight on interatomic interactions at the
surface. The observed composition modulation could be accounted for
quantitatively by the Defay-Prigogine and Strohl-King multilayer
extensions of the single-layer Gibbs model, revealing a near-surface
domination of the attractive Bi-Sn interaction over the entropy.
\end{abstract}



\maketitle


The widely-accepted Gibbs adsorption rule \cite{Gibbs28} predicts
the surface segregation of the lower surface energy component of a
binary mixture. Liquid metals are ideal systems for studying Gibbs
adsorption due to the nearly spherical shape of interacting
particles, relative simplicity of the short-range interactions and
the availability of bulk thermodynamic data for many binary alloys.
While certain aspects of Gibbs theory can be tested through
macroscopic measurements of surface tension or adsorption isotherms,
very few direct measurements of the atomic-scale composition
profiles of the liquid-vapor interface were reported \cite{Regan97,
Dimasi00b, Dimasi01}. In addition to fundamental questions related
to surface thermodynamics of binary liquids, BiSn-based alloys have
been widely studied as substitutes for Pb-based low-melting solders
\cite{Tu04}. Thus, understanding their wetting, spreading, alloying,
reactivity and other surface-related properties is of great
practical importance. Moreover, interfacial phenomena dominate the
properties of the increasingly important class of nanoscale
materials, as demonstrated recently in studies of the liquid-solid
phase stability of nanometer-sized BiSn particles \cite{Lee04}.

Synchrotron-based x-ray reflectivity (XR) can measure the
surface-normal density profile of a liquid with
{\AA}ngstr\"{o}m-scale resolution. Over the last decade XR revealed
the long-predicted surface-induced atomic layering at the
liquid-vapor interface for a number of elemental liquid metals
\cite{Magnussen95,Regan95,Tostmann99,Shpyrko03,Shpyrko04b}.
\textit{Resonant} XR near an absorption edge resolved the density
profile of each component in GaIn \cite{Regan97}, HgAu
\cite{Dimasi00b} and BiIn \cite{Dimasi01} liquid binary alloys. The
enhancement of the concentration of the low-surface-tension
component was invariably found to be confined to the topmost surface
monolayer, with subsequent layers having the composition of the
bulk, in accord with the simplest, and widely used, interpretation
of the Gibbs rule. By contrast, we find here  an atomic-scale phase
separation extending over at least three atomic layers. This is
unexpected, considering the near-perfect-solution nature of the
Bi$_{43}$Sn$_{57}$ alloy in the bulk \cite{Asryan04,Cho97}.

A liquid Bi$_{43}$Sn$_{57}$  sample (99.99$\%$ purity, Alfa Aesar)
was prepared under UHV conditions (P$<10^{-9}$~Torr). Atomically
clean liquid surfaces were obtained by mechanical scraping and
Ar$^{+}$ ion sputtering, as described previously \cite{Huber02,
Shpyrko03, Shpyrko04}. Measurements were done using the liquid
surface diffractometer at the ChemMatCARS beamline, Advanced Photon
Source, Argonne National Laboratory at a sample temperature of
$T=142~^{\circ}$C, $4~^{\circ}$C above the Bi$_{43}$Sn$_{57}$
alloy's eutectic temperature, $T_e=138~^{\circ}$C.

The reflected intensity fraction, $R(q_z)$, of an x-ray beam
impinging on a structured liquid surface at a grazing angle
$\alpha$, is given by the Born approximation as:
\begin{equation}
R(q_z) = R_F(q_z) \cdot  | \Phi (q_z) | ^2 \cdot CW(q_z)
\label{eq:rrf}
\end{equation}
where $q_z=(4\pi/\lambda)\sin \alpha$, $\lambda$ is the x-ray
wavelength, $R_{F}(q_z)$ is the Fresnel XR of an ideally abrupt and
flat surface, $CW(q_z)$ is due to thermal surface capillary waves
\cite{Tostmann99, Shpyrko03}, and the surface's structure factor is
\cite{AlsNielsen}:
\begin{equation}
\Phi (q_z) = \frac{1}{\rho_{\infty}} \int dz\frac{d \langle \rho(z)
\rangle }{dz} \exp(\imath q_z z). \label{eq:structure}
\end{equation}
Here $z$ is the surface-normal axis, $\rho_{\infty}$ and $\rho(z)$
are the bulk and surface electron densities, respectively, and
$\langle..\rangle$ denotes surface-parallel averaging. As $R_F$ is a
universal function depending only on the known critical angle for
total external reflection, and $CW(q_z)$ is known accurately from
capillary wave theory, the intrinsic density profile, $\langle
\rho(z) \rangle$, is obtained by computer fitting the measured
$R(q_z)$ by a physically motivated model described below
\cite{Shpyrko03}.

\begin{figure}[tbp]
\vspace{-0mm} \centering
\includegraphics[angle=0,width=0.9\columnwidth]{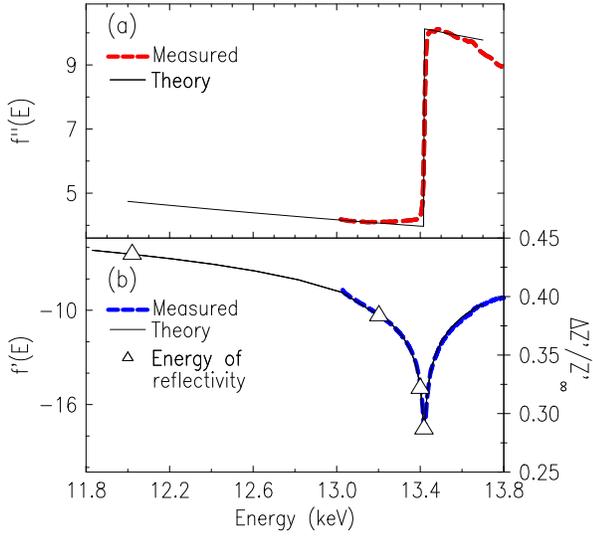}
\vspace{0mm} \caption{Dispersion corrections (a) $f'(E)$ and (b)
$f''(E)$ of Bi near the L3 absorption edge at $E_{L3}=13.418$~keV.
The right scale of (b) is the electron density contrast $\Delta
Z'/{Z'_{\infty}}=(Z'_{Bi}-Z'_{Sn})/Z'_\infty$.
}\label{fig:bisn_fig1}
\end{figure}
The (forward) atomic scattering factor of a $Z$-electron atom varies
with energy as \cite{AlsNielsen}: $Z'=Z+f'(E)-if''(E)$, where
$f'(E)$ and $f''(E)=\mu (E) \lambda/ (4\pi) $, are the real and
imaginary parts of the dispersion correction, and $\mu (E) $ is the
photoelectric absorption coefficient.

The effect of $f''$ on the analysis can be neglected and the changes
in $f'$ are significant only near an absorption edge.
Fig.~\ref{fig:bisn_fig1}(a) shows $f''(E)$ near the Bi L3 edge as
obtained from an absorption measurement in a Bi foil.
Fig.~\ref{fig:bisn_fig1}(b) is the corresponding $f'(E)$, calculated
from $f''(E)$ using the Kramers-Kronig relation \cite{Chooch}. Both
agree well with theory \cite{Merritt03,ifeffit}. The composition
dependence of $\langle \rho(z) \rangle$ was obtained by fitting the
measured XR by the distorted crystal (DC) model for a layered liquid
surface \cite{Magnussen95,Regan95}:
\begin{equation}~
\frac{\langle\rho(z)\rangle}{\rho_{\infty}}=\sum_{n=1}^{\infty}
\frac{e^{-(z-nd)^2/\sigma_n^2}}{\sqrt{2\pi}\sigma_n/d}
\Bigl(1+\delta_n \frac {
Z'_{Bi}-Z'_{Sn}}{Z'_\infty}\Bigr)\frac{c_n}{c}\label{eq:dist_cryst}
\end{equation}
The progressive increase in the Gaussian width parameter
$\sigma_n^2={\sigma_0^2+(n-1)\overline{\sigma}^2}$ with increasing
layer number $n$ describes the layering amplitude's decay below the
surface \cite{Regan95}. The layer spacing $d$ is kept constant in
this model due to similarity in size between Bi and Sn atoms. The
bulk's average effective electron number per atom is
$Z'_\infty=xZ'_{Bi}+(1-x)Z'_{Sn}$, and $\delta_n=x'_n-x$ is
difference in the Bi fraction between the $n$-th layer, $x'_n$, and
the bulk, $x$. The corresponding atomic densities, $c_n$ and $c$,
are determined from the atomic volumes $v_{Bi}$ and $v_{Sn}$: $c_n
x'_n v_{Bi}+c_n(1-x'_n)v_{Sn}=1$. The contrast,
$(Z'_{Bi}-Z'_{Sn})/Z'_\infty$, varies strongly near the edge due to
the variation of $Z'_{Bi}$: from $0.43$ at E=12.00~keV to $0.27$ at
E=13.418~keV (right axis in Fig.~\ref{fig:bisn_fig1}). This is the
basis for the resonant XR method which allows to separate out the
density profiles of the two species \cite{Dimasi01, Shpyrko04}.

\begin{figure}[tbp]
\centering
\includegraphics[angle=0,width=0.9\columnwidth]{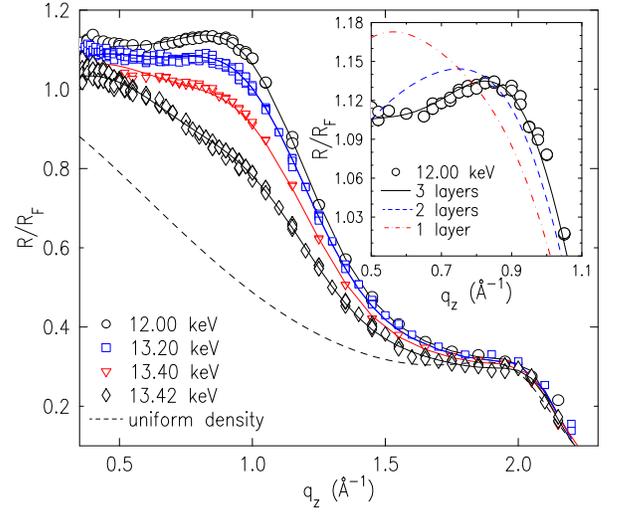}
\caption{XR measured at the indicated energies, with fits by the
three-layer model (lines). The dashed line is the XR of a
uniform-composition surface. Inset: The E=12.00~keV measured $R/R_F$
with fits by the three models discussed in the text (lines). Error
bars are smaller than the symbols' size.} \label{fig:bisn_fig2}
\end{figure}

Fig.~\ref{fig:bisn_fig2} shows Fresnel-normalized XRs
$R(q_z)/R_F(q_z)$ measured near the Bi L3 edge at the four energies
marked by triangles in Fig.~\ref{fig:bisn_fig1}(b). The dashed line
is calculated from the DC model for a layered interface assuming a
\emph{uniform} composition ($\delta_n=0$). The strong enhancement of
the measured $R(q_z)/R_F(q_z)$ over this line, evidenced by the peak
at $q_z \simeq 1.0~${\AA}$^{-1}$, and the strong energy-dependence
of the low-$q_z$ reflectivity, clearly indicate a significant
surface segregation of Bi, and its variation with $z$.

Three fits of the data by the DC model, Eq.~\ref{eq:dist_cryst},
were carried out, assuming that only one ($\delta_1 \neq 0$,
$\delta_{n \geq 2}=0$), two ($\delta_{1,2} \neq 0$, $\delta_{n \geq
3}=0$), or three ($\delta_{1,2,3} \neq 0$, $\delta_{n \geq 4}=0$)
surface layers deviate from the bulk composition. All fits employed
$d=2.90~${\AA}, $\sigma_0=0.30~${\AA} and
$\overline{\sigma}=0.57~${\AA}, derived from the energy-independent
position, shape and intensity of the layering peak at
$q_z=2.0~${\AA}$^{-1}$. The measured $R(q_z)/R_F(q_z)$\ of all four
energies were fitted simultaneously, using the experimentally
determined $f'(E)$.

Fig.~\ref{fig:bisn_fig2} exhibits an excellent agreement of the
three-layer model (solid lines) with the measurements, but a very
poor agreement for the one- and two-layer models (inset).
Table~\ref{tab:bisn_tab1} lists the  best-fit values of $x'_n$ and
$\delta_n^{Fit}=x'_n-x$  and the corresponding 95\% non-linear
confidence intervals $Y(x'_n)$ and $Y(\delta_n^{Fit})$ determined
from a six-parameter support plane analysis \cite{footnote}. The
most striking result is the non-monotonic deviation $\delta_n$ of Bi
from the 43\%\ bulk value, showing an enhanced composition of 96\%\
and 53\% in the first and third layers, and depletion down to 25\%
in the second layer (see Fig. 3). Beyond the third layer entropy
effects dominate the Gibbs adsorption and the layer and bulk
concentrations can not be distinguished. While demixing has not been
previously reported in liquid alloys, similar decaying oscillatory
composition profiles were discovered in several \emph{crystalline}
alloys such as Cu$_3$Au \cite{Reichert95}. However, the properties,
formation mechanism, and strong temperature dependence of the
composition modulations in Cu$_3$Au alloys were found to be
intimately related to, and largely dominated by, the long-range fcc
order in the bulk crystal, and the severe packing strains resulting
from the big mismatch in the atomic diameter of the two components.
As none of these exist in our liquid alloy, surface-induced
segregation and the Gibbs rule can be studied in a pure
short-range-order interaction context, free from the complicating
influence, or even dominance, of other effects. We now compare our
experimental observations with theory.

\begin{figure}[tbp] \vspace{-0mm} \centering
\includegraphics[angle=0,width=0.9\columnwidth]{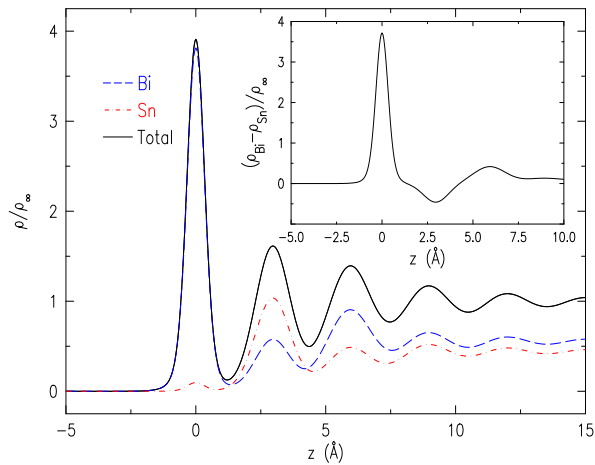}
\caption{Electron density profiles as derived from the fits to the
reflectivities shown in Fig.~\ref{fig:bisn_fig2}. Inset: the
bulk-normalized differences in electron density of Bi and Sn,
$(\rho_{Bi}-\rho_{Sn})/\rho_\infty$.} \label{fig:bisn_fig3}
\end{figure}

Guggenheim's \cite{Gugg45} application of Gibbs theory
\cite{Gibbs28} to regular solutions assumes the surface segregation
to be restricted to a single surface monolayer.  Assuming $p$
nearest neighbors for each bulk atom in a layered lattice model,
$lp$ are within, and $mp$ are in the adjacent, layers. For a
close-packed lattice, for example, $p=12$, $l=0.5$ and $m=0.25$. The
surface tension of the regular solution, $\gamma_{AB}$, follows from
those of the pure components, $\gamma_A$ and $\gamma_B$, as
 \cite{Gugg45}:
\begin{eqnarray}
\gamma_{AB}=\gamma_{B}+\frac{kT}{a_B}\ln\Bigl(\frac{1-x'}{1-x}\Bigr)+\frac{\omega}{a_B}[lx'^2-(l+m)x^2]~~\label{eq:gibbs}\\
=\gamma_{A}+\frac{kT}{a_A}\ln\Bigl(\frac{x'}{x}\Bigr)+
\frac{\omega}{a_A}[l(1-x')^2-(l+m)(1-x)^2].\nonumber
\end{eqnarray}
Here, $x$ and $(1-x)$ are the bulk concentrations of atoms A (Bi)
and B (Sn), while $x' \neq x$ and $(1-x')$ are the corresponding
surface concentrations, $a_A$ and $a_B$ are the two atomic areas,
and $\omega=2\omega_{AB}-\omega_{AA}-\omega_{BB}$ is the interaction
parameter, defined by the A-B, A-A and B-B atomic interaction
energies. Extrapolated down to $T=142~^{\circ}$C, $\gamma_{Bi}=398~$
mN/m and $\gamma_{Sn}=567~$mN/m, while $a_{Bi}$\ and $a_{Sn}$ are
calculated from the atomic radii $r_{Bi}= 1.70$~{\AA}\ and $r_{Sn}=
1.62$~{\AA}\ assuming hexagonal close packing \cite{CRC}. Treating
Bi$_{43}$Sn$_{57}$ as a perfect solution ($\omega/kT = 0$), the
Gibbs theorem, Eq.~\ref{eq:gibbs}, yields $\gamma_{AB}=444~$mN/m and
$x'=0.904$, below the experimental value $x'_1$(Bi) in Table
\ref{tab:bisn_tab1}. However, assuming a regular solution behavior
with $\omega/kT=1$ yields $\gamma_{AB}=432~$mN/m, and $x'=0.941$,
which agrees very well with the experimentally derived $x'_1({\rm
Bi})$ in Table~\ref{tab:bisn_tab1}. Both $\gamma_{AB}$ agree well
with experiment and theory \cite{Yoon99}. Note that $\gamma_{AB}$
and $x'$ are only weakly dependent on $\omega/kT$ due primarily to
the logarithmic functional behavior and large surface tension
difference of the two components,
$a_{Sn}\gamma_{Sn}-a_{Bi}\gamma_{Bi}\approx 2~kT$. This introduces a
large uncertainty of $\omega/kT$ calculated from measurements of
surface tension or surface monolayer composition. Resonant XR
measurements of sub-surface layer composition therefore present a
unique opportunity to probe the nature of atomic interactions at the
surface.
\begin{table}[tcp]
\squeezetable
\begin{ruledtabular}
\begin{tabular}{ccccccc}
        $n$ & $x'_n (Bi)$ &      $Y(x'_n)$   & $\delta_n^{Fit}$ &   $Y(\delta_n^{Fit})$   &   $\delta_n^{DP}$  &   $\delta_n^{SK}$   \\
\hline
         1 &      0.96 &      [0.94,~0.99] &      0.53  &  [0.51,~0.56]     &       0.47    &    0.51       \\
         2 &      0.25 &      [0.18,~0.27] &     -0.18  &  [-0.25,~-0.16]   &       -0.23   &   -0.25       \\
         3 &      0.53 &      [0.50,~0.56] &      0.10  &  [0.07,~0.13]     &       0.12    &    0.05       \\
         4 &      0.43 &       -           &          0 &       -           &       -0.06   &   -0.01       \\
\end{tabular}
\end{ruledtabular} \caption{Density model parameters $x'_n$ and $\delta_n^{Fit}=x'_n-x$,  and confidence intervals $Y(x'_n)$
and $Y(\delta_n^{Fit})$ obtained from  the three-layer model fits
compared to theoretical $\delta_n^{DP}$ and $\delta_n^{SK}$ derived
from the Defay-Prigogine and Strohl-King models.}
\label{tab:bisn_tab1}
\end{table}

In spite of the good agreement above,  confining the surface excess
to a single monolayer is correct for perfect solutions only, but not
for our case of a regular solution, as Defay and Prigogine
\cite{Defay50} point out. They provide a correction for regular
solutions, where the surface excess extends over two layers,
 the $\gamma_{AB}$ values above do not change significantly
and the layers' $\delta_n$ are related by:
\begin{equation}
\ln\frac{1+\delta_2/x} {1-\delta_2/(1-x)} -\frac{2 \omega
}{kT}\delta_2-\frac{2 \omega m}{kT}(\delta_1-2\delta_2)=0.
\label{eq:prigogine}
\end{equation}
Expanding Eq.~\ref{eq:prigogine} to first order in $\delta_2$:
\begin{equation}
\delta_2=\frac{{2 \omega m}x(1-x)\delta_1}{kT-2\omega lx(1-x)}.
\label{eq:prigogine2}
\end{equation}
For nearly perfect solutions ($\omega /kT \ll 1$)
Eq.~\ref{eq:prigogine} yields a negligible  $\delta_2$: $0< \delta_2
\ll \delta_1$. For $\omega /kT \gtrsim 1$, however, $\delta_2$ and
$\delta_1$ are of opposite signs and $|\delta_2|$ may become
comparable to $|\delta_1|$. This prediction is qualitatively
consistent with the demixing observed here. For example, when
$\omega /kT \gg 1$, Eq.~\ref{eq:prigogine2} can be simplified
further: $\delta_2=-(m/l)\delta_1$. For Bi$_{43}$Sn$_{57}$,
$m/l\approx0.5$ and the Gibbs-predicted $x'_1=0.90$ (or
$\delta_1=0.47$) yields $\delta_2=-0.23$, $\delta_3=0.12$ and
$\delta_4=-0.06$ \cite{rec}. These values, shown as $\delta_n^{DP}$
in Table~\ref{tab:bisn_tab1}, agree well with $\delta_n^{Fit}$
obtained from the three-layer model fits. The smallest value of the
interaction parameter $\omega/ kT$ for which satisfactory agreement
with the Defay-Prigogine model could be obtained (by treating $m$ as
an adjustable parameter) is $\omega/ kT = 2.3$.  Strohl and King
\cite{Strohl89} suggest a multilayer, multicomponent model, where no
expansion is used, and $x'_n$ are obtained iteratively, until
convergence to a self-consistent composition profile is reached. A
good agreement of this theory with our BiSn measurements is obtained
when $\omega/ kT =1.0-1.7$. Typical $\delta_n^{SK}$ values are
listed in Table~\ref{tab:bisn_tab1}. As observed, the Strohl-King
model provides composition profiles very similar to those of the
Defay-Prigogine model, albeit with slightly different $\delta_n$
values, thus supporting our overall conclusions.

Theoretically, $\omega$ and the enthalpy of mixing, $\Delta H_m$,
are related by $\omega=\Delta H_m/[x(1-x)]$. In practice, however,
bulk thermodynamic quantities were often found to yield inaccurate
values for surface quantities. For example, organic \cite{Gaines69}
and metallic \cite{Hoar57} mixtures exhibit significant
disagreements between $\omega$ values  derived empirically from
surface tension measurements and from bulk calorimetry. For BiSn,
reported values of $\Delta H_m$ range from endothermic values of 80
to 140 J/mol \cite{Asryan04} to an exothermic value of -180 J/mol
\cite{Cho97}. These values lead to $|\omega/kT|<0.2$, i.e. an almost
perfect solution, and an insignificant $|\delta_2|<0.01$. On the
other hand, the value of $\omega/ kT \approx 10$ that we previously
found necessary to account for the observed 35\%\ Bi concentration
enhancement at the surface monolayer at the BiIn eutectic is of the
same order of magnitude as the value we find necessary to account
for the present observation of surface segregation in BiSn, $\omega/
kT \approx 1.0-2.3 $. Unfortunately we do not have an explanation
for the origin of the discrepancy in the values of $\omega/ kT$ and
this suggests an urgent need for both further theoretical studies of
surface demixing as well as experimental investigations of similar
effects in other binary alloys. In particular, the BiSn system
appears to be the only liquid alloy for which clear evidence for
multilayer surface demixing has been found. The case for new studies
is strongly reinforced by the existence of a growing class of
surface-induced ordering phenomena that have been observed in
metallic liquids. In addition to the surface demixing reported here,
these include layering
\cite{Magnussen95,Regan95,Tostmann99,Shpyrko03, Shpyrko04b},
relaxation \cite{Shpyrko04b}, segregation
\cite{Regan97,Dimasi00b,Dimasi01, rice93}, wetting transitions
\cite{Huber02,wynblatt01}, and surface freezing \cite{Shpyrko05}.
Finally, there is a basic unresolved question of whether the
surfaces of liquid metals are fundamentally different from those of
non-metallic liquids \cite{Layer}.

This work has been supported by the U.S. DOE grants No.
DE-FG02-88-ER45379, DE-AC02-98CH10886 and the U.S.-Israel Binational
Science Foundation, Jerusalem. We gratefully acknowledge useful
discussions with E. Sloutskin at Bar-Ilan as well as assistance from
T. Graber, D. Schultz and J. Gebhardt at ChemMatCARS Sector 15,
principally supported by the NSF/DOE grant No. CHE0087817. The
Advanced Photon Source is supported by the U.S. DOE contract No.
W-31-109-Eng-38.

\bibliographystyle{prsty}

\end{document}